\documentclass[twocolumn,amsmath,amssymb]{snp}
\usepackage{graphicx}
\usepackage{dcolumn}
\usepackage{bm}
\begin{document}

\title{{\Large N/Z dependence of balance energy as a probe of symmetry energy in heavy-ion collisions}}

\author{\large Sakshi Gautam$^1$}
\author{\large Aman D. Sood$^2$}
\author{\large Rajeev K. Puri$^1$}
\email{rkpuri@pu.ac.in}
\affiliation{$^1$Department of Physics, Panjab University,
Chandigarh - 160014, INDIA} \affiliation{$^2$SUBATECH, Laboratoire de Physique Subatomique et des Technologies Associ\'{e}es, Universit\'{e} de Nantes - IN2P3/CNRS - EMN \\
4 rue Alfred Kastler, F-44072 Nantes, France} \maketitle

\section*{Introduction}
The equation of state (EOS) of symmetric nuclear matter is well
understood due to extensive efforts of nuclear physics community
for the past few decades. Nowadays, the nuclear EOS of asymmetric
nuclear matter has attracted lot of attention. The EOS of
asymmetric nuclear matter can be described approximately by:
\begin{equation}
E(\rho, \delta) = E_{0}(\rho, \delta=0)+E_{sym}(\rho)\delta^{2}
\end{equation}
where
 $\delta$ = $\frac{\rho_{n}-\rho_{p}}{\rho_{n}+\rho_{p}}$ is isospin asymmetry,
  E$_{0}$($\rho$, $\delta$) is the energy of pure symmetric nuclear matter and E$_{sym}$($\rho$)
  is the symmetry energy with E$_{sym}$($\rho$$_{0}$) = 32 MeV is the symmetry energy at normal
  nuclear matter density. The symmetry energy is E($\rho$,1) - E$_{0}$($\rho$,0), ie. difference
  between energy per nucleon between pure neutron matter and symmetric nuclear matter.
  Symmetry energy is important not only for the nuclear physics community
  as it sheds light on the structure of radioactive nuclei, reaction dynamics induced
by the rare isotopes but also to the astrophysicists as it acts as
a probe for understanding the evolution of massive stars and the
supernova explosion \cite{marr10,kubis07}.
 Experimentally, symmetry energy is not a directly
measurable quantity and has to be extracted from the observables
which are related to the symmetry energy. Therefore, a lot of
observables have been proposed in recent past which act as a probe
of symmetry energy and its density dependence \cite{symm}.
Collective transverse in-plane flow has been studied for past
three decades and also being used to explore the isospin
dependence of EOS \cite{pak,gaum10}. The energy dependence of
collective flow led its disappearance at balance energy E$_{bal}$.
In the present study, we propose N/Z dependence of E$_{bal}$ as a
probe of symmetry energy and its density dependence. The present
study is carried within the framework of isospin-dependent quantum
molecular dynamics (IQMD) model \cite{hart98}.


\section*{Results and Discussions}
\begin{figure}[!t]
\centering \vskip 0cm
\includegraphics[angle=0,width=5cm]{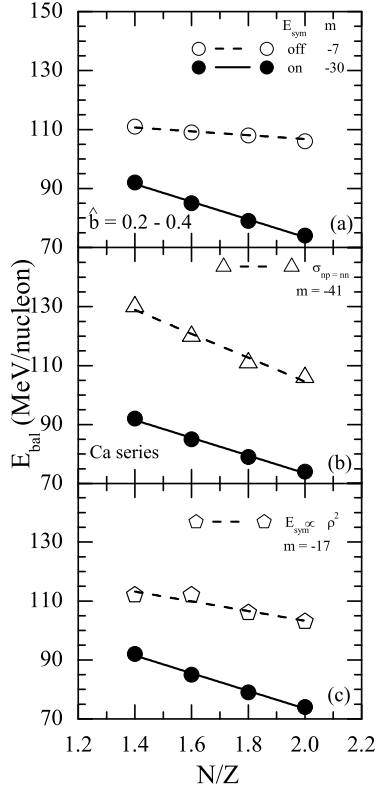}
\vskip 0cm \caption{N/Z dependence of E$_{bal}$ with symmetry
energy on and off, (b) isospin-dependent and independent nn cross
section, and (c) E$_{sym}$ $\varpropto$ $\rho$$^{2}$. Various
symbols and lines are explained in the text.}\label{fig1}
\end{figure}
We simulate the reactions of Ca+Ca having
 N/Z varying from 1.4 to 2.0 in small steps of 0.2, viz. Ca$^{48}$+Ca$^{48}$, Ca$^{52}$+Ca$^{52}$, Ca$^{56}$+Ca$^{56}$, and Ca$^{60}$+Ca$^{60}$ at impact
parameter of $\hat{b}$=0.2-0.4. We use a soft EOS along with
anisotropic standard isospin- and energy-dependent nn cross
section $\sigma$ = 0.8 $\sigma$$_{NN}$$^{free}$ \cite{gaum10}. The
cross sections for neutron-neutron collisions are assumed to be
equal to the proton-proton cross sections. Fig. 1 (a) displays the
N/Z dependence of E$_{bal}$ for Ca+Ca (solid circles) throughout
the isotopic series. We find that E$_{bal}$ decreases with
increase in N/Z due to increased role of symmetry energy in
systems with higher N/Z and follows a linear behaviour with N/Z.
Isospin effects in collective transverse in-plane flow is due to
competition between symmetry energy and isospin-dependent nn cross
section. To see the dominance among these two, we switch off the
symmetry energy and also make the nn cross section isospin
independent, respectively. The preliminary results are displayed
in Fig. 1 (a) (open circles) and (b) (open triangles) with no
symmetry energy and isospin independent cross section,
respectively. From the figure, we find that E$_{bal}$ increases
when we switch off the symmetry energy. This is due to the fact
that symmetry energy contributes to repulsive potential and in the
absence of symmetry potential, the collective transverse in-plane
flow decreases leading to enhancement of E$_{bal}$. Moreover, the
increase in the E$_{bal}$ is more for higher N/Z system which is
due to the increased role of symmetry energy in respective system.
The behavior of E$_{bal}$ with symmetry energy off is still linear
but with reduced slope from -30 to -7. This indicates that the N/Z
dependence of E$_{bal}$ is sensitive to symmetry energy. To
further strengthen our point, fig. 1 (b) shows E$_{bal}$ with
isospin independent nn cross section ,i.e. $\sigma$$_{np}$ =
$\sigma$$_{nn}$ or $\sigma$$_{pp}$. E$_{bal}$ with isospin
independent cross section increases which is due to the fact that
isospin-dependent cross section $\sigma$$_{np}$ = 3$\sigma$$_{nn}$
or $\sigma$$_{pp}$, but when we make the cross section isospin
independent, the effective total cross section decreases leading
to less transverse in-plane flow and hence enhancement of
E$_{bal}$. From the figure, we also see that change in the
E$_{bal}$ is more compared to one in fig. 1 (a) pointing towards
increasing role of isospin dependence of nn cross section in
E$_{bal}$. But N/Z dependence of E$_{bal}$ is still linear with
almost the same slope as with isospin-dependent nn cross section,
thus indicating almost negligible role of isospin dependence of
cross section as far as the N/Z dependence of E$_{bal}$ is
concerned. This clearly demonstrates the dominance of symmetry
energy in N/Z dependence of E$_{bal}$. To see the effect of
density dependence of symmetry energy, fig. 1(c) (open pentagons)
displays E$_{bal}$ with symmetry energy $\varpropto$ $\rho^{2}$.
From the figure, we see that with E$_{sym}$ $\varpropto$
$\rho^{2}$, E$_{bal}$ increases and its N/Z dependence is linear
with reduced slope of -17 as compared to one with linear density
dependence (solid circles) indicating that N/Z dependence of
E$_{bal}$ can act as a probe of density dependence of symmetry
energy as well.

\section*{Acknowledgments}
 This work is supported by Indo-French project
no. 4104-1.


\end{document}